\documentclass[conference]{IEEEtran}
\IEEEoverridecommandlockouts
\usepackage[frozencache,cachedir=.]{minted}
\usepackage[hyphens]{url}
\usepackage{cite}
\usepackage{amsmath,amssymb,amsfonts}
\usepackage{algorithmic}
\usepackage{graphicx}
\usepackage{textcomp}
\usepackage{xcolor}

\makeatletter
\let\@float@c@listing\@caption
\makeatother

\def\BibTeX{{\rm B\kern-.05em{\sc i\kern-.025em b}\kern-.08em
    T\kern-.1667em\lower.7ex\hbox{E}\kern-.125emX}}
    
\usepackage{graphicx,calc}
\newlength\myheight
\newlength\mydepth
\settototalheight\myheight{Xygp}
\newcommand*\inlinegraphics[1]{%
  \settototalheight\myheight{Xygp}%
  \settodepth\mydepth{Xygp}%
  \raisebox{-\mydepth}{\includegraphics[height=\myheight]{#1}}%
}

\begin{document}

\title{Rapid Development of Compositional AI\\

\author{\IEEEauthorblockN{Lee Martie, Jessie Rosenberg, Veronique Demers, Gaoyuan Zhang, Onkar Bhardwaj, John Henning, \\Aditya Prasad, Matt Stallone, Ja Young Lee, Lucy Yip, Damilola Adesina, \\Elahe Paikari, Oscar Resendiz, Sarah Shaw, David Cox}
\IEEEauthorblockA{MIT-IBM Watson AI Lab, IBM Research \\
Cambridge, Massachusetts 02142 USA \\
\{lee.martie, jcrosenb, vdemers, gaoyuan.zhang, onkarbhardwaj, john.l.henning, aditya.prasad, \\mstallone, ja.young.lee, lucy.yip, dami.adesina, epaikari, oscar.resendiz, sarah.y.shaw, david.d.cox\}@ibm.com
}
}


}

\author{\IEEEauthorblockN{\\\textcolor{white}{.}\\\textcolor{white}{.}\\\textcolor{white}{.}}}
\maketitle

\begin{abstract}
Compositional AI systems,
which combine multiple artificial intelligence components together with other
application components to solve a larger problem, have no
known pattern of development and are often approached in a
bespoke and ad hoc style. This makes development slower and
harder to reuse for future applications. To support the full rapid development cycle of compositional
AI applications, we have developed a novel framework called
\((Bee)*\) (written as a regular expression and pronounced as ``beestar''). We illustrate how \((Bee)*\) supports building integrated, scalable,
and interactive compositional AI applications with a simplified
developer experience.
\end{abstract}


\section{Introduction}
\label{sec:intro}

Delivering applications to users quickly is a cornerstone of agile development methodologies \cite{shore_art_2007}, allowing multiple iteration cycles to identify and incrementally solve critical stakeholder problems in response to working applications. However, rapidly developing software requires a developer to focus their effort on what is novel about an application, without distraction from secondary or accidental tasks \cite{brooks_no_1987}. 

Rapidly developing novel systems is difficult today as applications frequently incorporate various types of artificial intelligence (AI) methods such as neural networks, symbolic knowledge representations, reinforcement learning, and others. Compositional AI systems, which combine multiple AI components together with other application components to solve a larger problem, have no known pattern of development and are often approached in a bespoke and ad hoc style. This makes development slower and/or error prone as time needs to be spent on design or coding with no clear architecture, distracting the developer's attention from critical features.

Tools today support rapid development of AI programs to some extent. For example, Streamlit \cite{noauthor_streamlit_nodate} and Gradio \cite{team_gradio_nodate} offer a solution to rapidly create frontends by enabling the developer of a Python program (the most common language of AI) to automatically display static data in a web browser and interact with it through a variety of widgets. However, they do not solve several critical challenges in creating compositional AI applications \cite{noauthor_is_nodate, noauthor_how_2021}. They offer no support for \textbf{integrating} multiple AI components, \textbf{scaling} these components on the needed infrastructure, and building dynamically updating applications for rich AI and user \textbf{interaction}.

To support the full rapid development cycle of compositional AI applications, we have developed a novel framework called \((Bee)*\) (written as a regular expression and pronounced as ``beestar''). This framework enables the developer to declaratively build a representation of the entire application (specifying AI components, component and user interactions, visualizations, and desired scaling) in a graph structure that \((Bee)*\) interprets in order to automatically create and scale the application. This approach supports rapid application development by simplifying the developer's responsibility to deciding how to compose AI and GUI elements with declarations, leaving the task of operationally composing the application to \((Bee)*\). 

\begin{figure*}[!tp]
    \centering
    \includegraphics[width=0.8\textwidth]{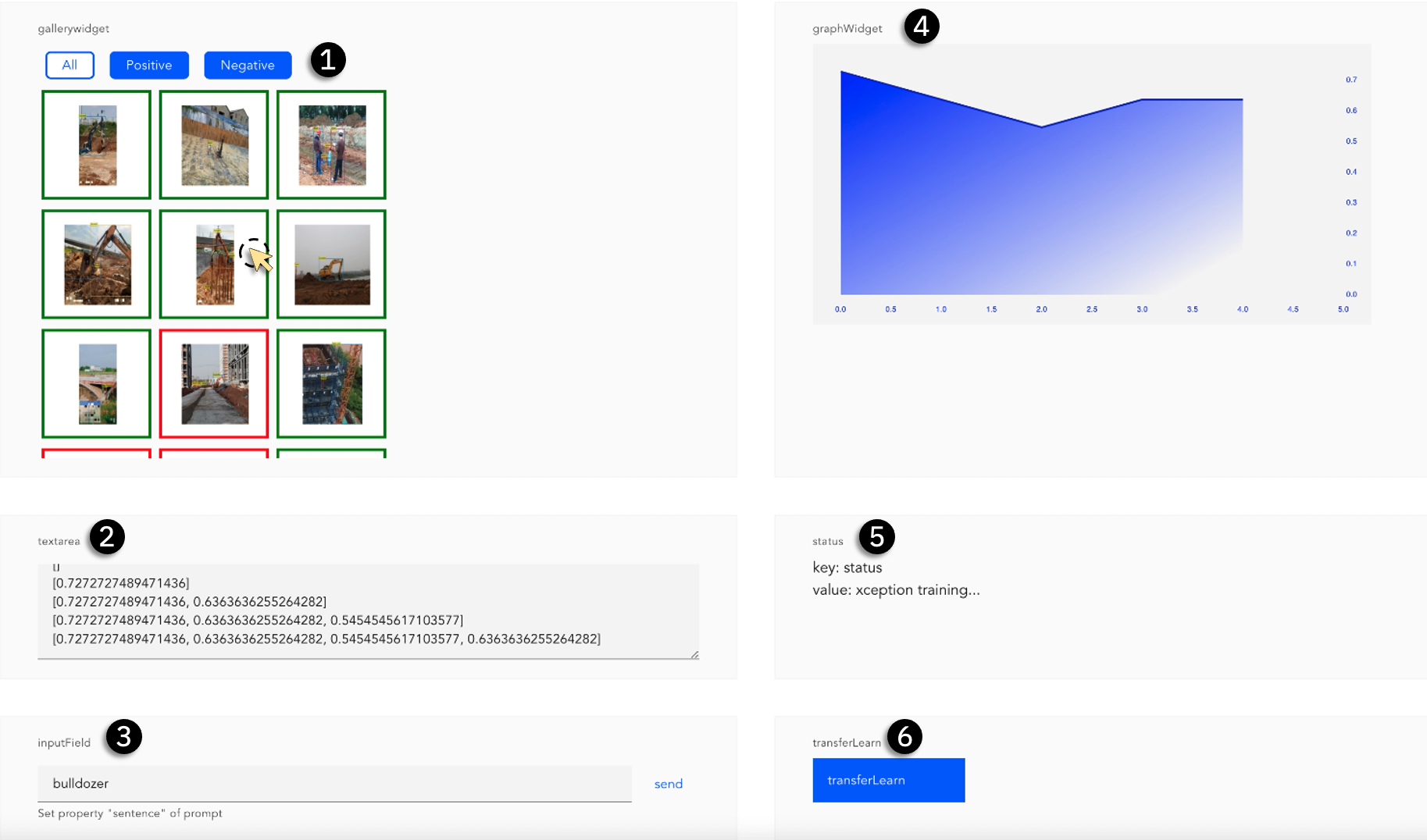}
    \caption{Dashboard Example.}
    \label{fig:dasboard}
\end{figure*}

To leverage the graph created by the developer and operate at scale, \((Bee)*\) follows an agent-oriented \cite{jennings_agent-oriented_1999,  hayes-roth_blackboard_1985} approach to programming, where \textit{agents} are the computational components of the program. \((Bee)*\) agents are independent run-times that can take messages to execute their code, coordinate with other agents, and interact with the user with \textit{widgets}, through the graph. As the entire specification of the application is on the graph, widgets and agents can update the application by updating the graph and the \((Bee)*\) runtime will dynamically update the application. Further, the graph supports  meta-programming for very dynamic behavior, where agents and widgets use it to change source code at runtime.

In Section \ref{sec:Graph_Application_Representation}, we use a compositional AI application built using \((Bee)*\) to show how \((Bee)*\) supports \textbf{integrated}, \textbf{scalable}, and \textbf{interactive} compositional AI applications with a simplified developer experience. 
In Section \ref{sec:Evaluation}, we discuss the feasibility of building with \((Bee)*\). In Section \ref{sec:Related_Work} we discuss related work. In Section \ref{sec:Future_Plans} and \ref{sec:Conclusion} we discuss future plans and conclude.

\section{\((Bee)*\) System}
\label{sec:Graph_Application_Representation}

\((Bee)*\) uses a graph representation to create and dynamically update applications for compositional AI.  Figure \ref{fig:dasboard} shows a GUI (called a Dashboard) of a compositional AI application created with such a graph. Figure \ref{fig:dasboard}, \inlinegraphics{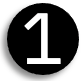}, shows a gallery of training data, where each image has been labeled as ``bulldozer'' (green frame) or not (red frame) by the AI model CLIP \cite{radford_learning_2021}, or by the user correcting CLIP by clicking on an image to toggle the label. The CLIP agent is run to label the images as ``bulldozer'' in response to the user typing this word into the prompt (Figure \ref{fig:dasboard}, \inlinegraphics{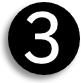}). Once the user is satisfied with the labels, the user clicks the TranserLearn button (Figure \ref{fig:dasboard}, \inlinegraphics{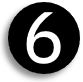}) to trigger another process that fine-tunes the Xception model \cite{chollet_xception_2017} on the labeled data so that it can predict bulldozer images (validation accuracy per epoch shown in \ref{fig:dasboard}, \inlinegraphics{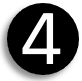}). Status (Figure \ref{fig:dasboard}, \inlinegraphics{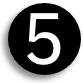}) and logging (Figure \ref{fig:dasboard}, \inlinegraphics{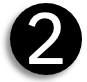}) widgets show updates from the AI components. 

\begin{figure*}[!tp]
    \centering
    \includegraphics[width=0.71\textwidth]{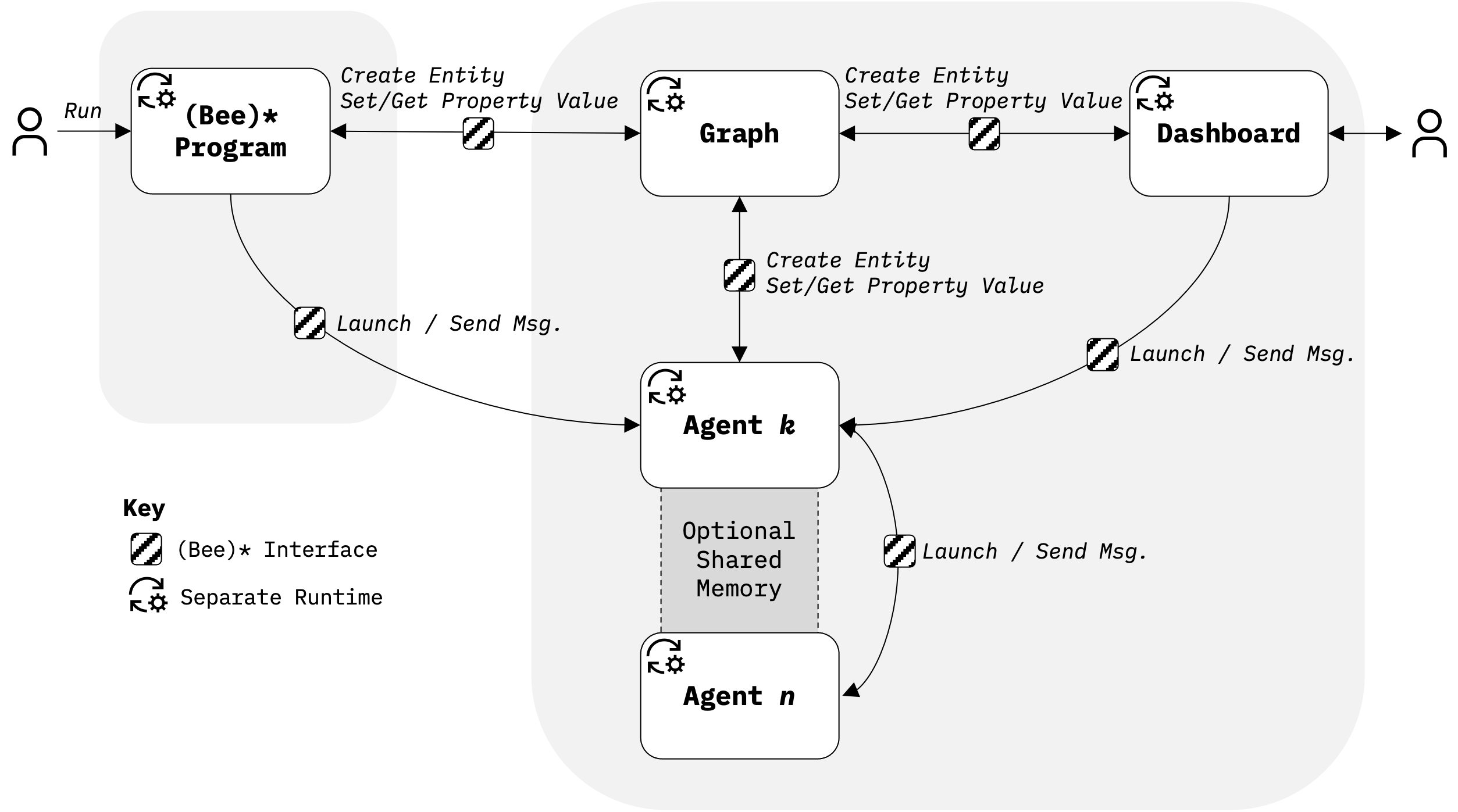}
    \caption{\((Bee)*\) Architecture.}
    \label{fig:arch}
\end{figure*}

To create the application in Figure \ref{fig:dasboard}, a graph is specified and executed by the \((Bee)*\) architecture (Figure \ref{fig:arch}). The major components are (1) \textbf{\((Bee)*\) program}, written by the developer, which declaratively specifies a graph, the (2) \textbf{\((Bee)*\) graph} itself (hosted on some graph database), (3) \textbf{agents} that are standalone run-times executing different algorithms/programs and interacting through the graph, and the (4) \textbf{Dashboard} that is populated with widgets, specified in the graph, that the user can, in turn, use to view/update the graph. 



\subsection{Graph Representation}
We describe a \((Bee)*\) graph through an example (Figure \ref{fig:graph}) used to create the application shown in Figure \ref{fig:dasboard}. Nodes in a graph are entities, which specify arbitrary concepts (similar to objects in object oriented programming \cite{booch_object_1991}) with properties and values. For example, in Figure \ref{fig:graph} we have an entity named \textit{Training Data} with property \textit{data} and value \textit{link}, where \textit{link} names where the value is stored (in this case we assume a path on disk). This entity will be used by the AI components (i.e., agents) for training and inference. Further, \((Bee)*\) provides a primitive typing system, so we know that whatever value \textit{data} is, it is an array. \((Bee)*\) knows the entity type of  \textit{Training Data} using the edge with name ``is a'' to the entity named Entity.

\subsubsection{Integration with Agents and AgentEntities}
In \((Bee)*\), all processes/algorithms/AI components are agents \cite{jennings_agent-oriented_1999}, where an agent is a standalone Python runtime that can take messages (using a remote procedure call server \cite{bloomer_power_1992}) and map messages to different behaviors. When an agent receives the message ``play'' it runs its assigned source code, ``stop'' will stop execution, and ``debug'' will run assigned source code with the Python debugger \cite{noauthor_pdb_nodate}. The agent itself is specified as an entity (i.e., AgentEntity) that \((Bee)*\) uses to start the agent and as a place to keep and update properties of the agent. The agent also uses its AgentEntity specification to locate its source code to execute on the message ``play''. By specifying the source code as a property of the agent, the agent is able to reflect on its own source code as data or other agent's source code to support meta-programming \cite{czarnecki_generative_2002}. Use cases for this include code injection for logging \cite{bergmans_aspect-oriented_1999}, code optimization \cite{aho_compilers_2006}, or genetic programming \cite{banzhaf_genetic_1998} for functional changes. Further, source code as a property supports the user to debug and change the agent's source code at runtime with widgets described in Section. \ref{sec:Dashboard_and_WidgetEntities}.

\begin{figure*}
    \centering
    \includegraphics[width=0.90\textwidth]{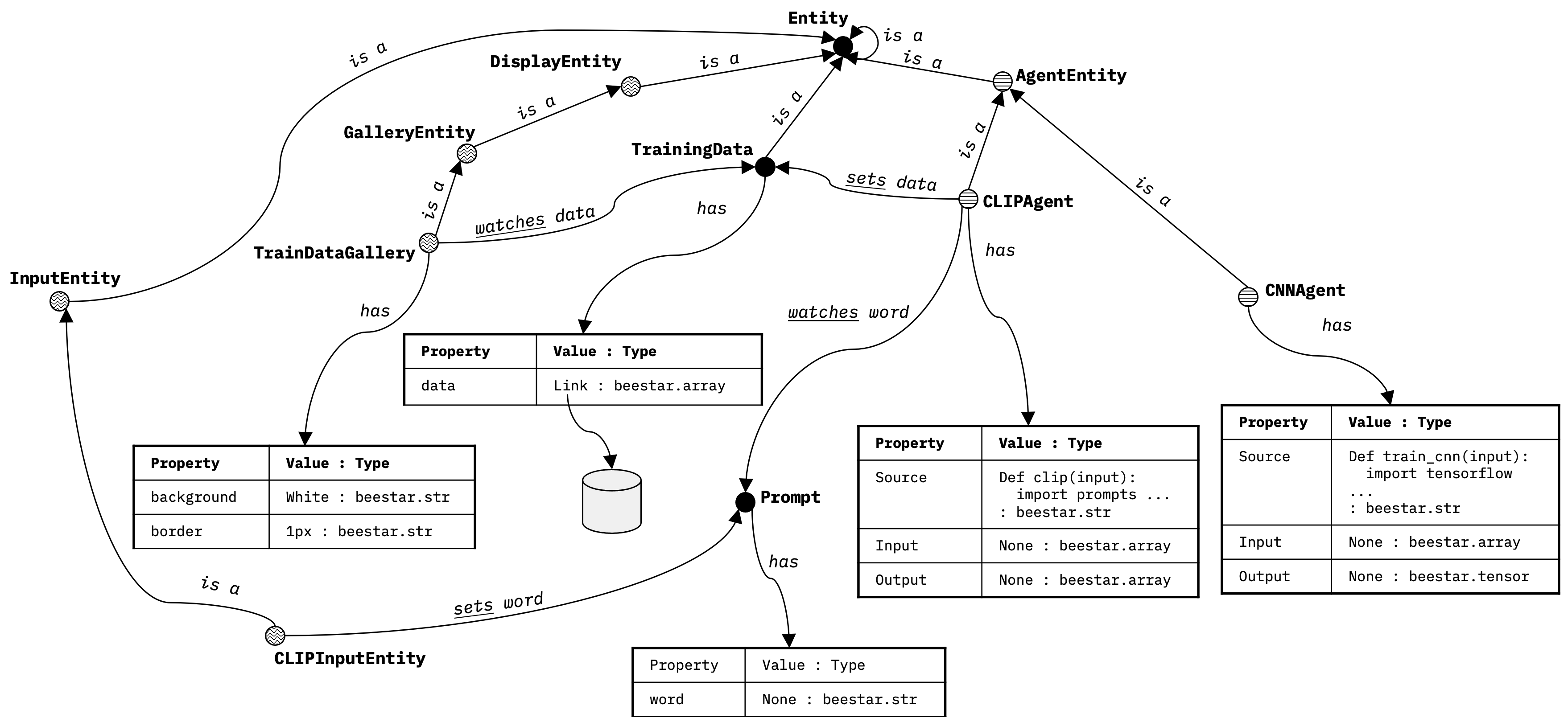}
    \caption{Part of Graph for Compositional AI Application Example.}
    \label{fig:graph}
\end{figure*}

AgentEntity inherits from Entity and has the properties \textit{source code} (defining a function), \textit{input} and \textit{output} that specify an agent that \((Bee)*\) can launch and run. In Figure \ref{fig:graph}, there are AgentEntities \textit{CLIPAgent} and \textit{CNNAgent}, where \textit{CLIPAgent} has \textit{source code} to run the CLIP model \cite{radford_learning_2021} to label arbitrary images. \textit{CNNAgent}'s source code trains a CNN model (Xception \cite{chollet_xception_2017}). When an agent executes its source code, it runs by passing the value of its \textit{input} property to the function and maps the return value to its \textit{output} property in its AgentEntity. In this way, the source code, input, and output are ``lifted out'' of the agent's runtime and usable and modifiable by any other agents at the graph level, where the agents could be running very different source code and living across different servers on a cluster. As such, agents are able to run code in different languages (with operating system calls) but \textbf{input and output are independent of their language, supporting integration across languages}. This is helpful in scenarios, for example, where we need to integrate probabilistic programming in Julia \cite{noauthor_julia_nodate} with a model in Python.

\subsubsection{Display Entities and Input Entities for User Interaction}
\label{sec:Dashboard_and_WidgetEntities}
\((Bee)*\) provides what is called the Dashboard to the user to interact with and view the state of the graph. Widgets on the Dashboard are specified by the developer and, just like agents, widgets are specified as Entities. The developer can specify DisplayEntities (widgets for viewing), InputEntities (widgets for changing properties on Entities), and ButtonEntites to send messages to agents. Custom widgets are possible by extending these types. Once these Entities are declared, the Dashboard finds them on the graph and populates its interface with them. In Figure \ref{fig:graph}, there is a \textit{TrainDataGallery} entity of type GalleryEntity that the Dashboard uses to setup the view of construction data in Figure \ref{fig:dasboard}, \inlinegraphics{figures/one.png}, and customizes it's background and border using the \textit{TrainDataGallery}'s properties. A \textit{GalleryEntity} can display any images on the graph by adding a relation specified by an edge (described in next section) from it to the data source, allowing for changes of data sources at runtime and dynamic behavior.

\subsubsection{Agent and User Interaction with Watches and Sets}
Any kind of DisplayEntity (e.g., GalleryEntity) displays data of interest by specifying an edge to the entity with the property of interest, where the label on the edge is ``watches $\langle property\ name \rangle$''. With this edge, the DisplayEntity is notified to update on any change of that property's value. For example, \textit{TrainDataGallery} displays the training data because it ``watches data'' of the \textit{TrainingData} Entity. When the data is updated, \((Bee)*\) notifies the \textit{TrainDataGallery} to also update. Similarly,  an agent is triggered to run its source code if some property it is interested in changes. In Figure \ref{fig:graph}, if the property \textit{word} of \textit{Prompt} is changed, then \textit{CLIPAgent} is triggered to run its source code because it is watching that property. When triggered, \textit{CLIPAgent}'s input property is set to the value of \textit{word} and passed as an argument to \textit{CLIPAgent}'s source code.

To support modularity, the output value of an agent can update the value of other properties with the \textit{sets} relationship. In our example, there is an edge from \textit{CLIPAgent} to \textit{Training Data} with label ``sets data'', which will set the value of the data property of \textit{Training Data} to the output of \textit{CLIPAgent}. Using this relation has the advantage that it decouples AgentEntities and DisplayEntities. For example, \textit{TrainDataGallery}, will update with new \textit{data} without being coupled to the \textit{CLIPAgent} agent. Further, InputEntities can also set values of properties. In Figure \ref{fig:graph}, we have an InputEntity, \textit{CLIPInputEntity} (displayed in Figure 1, \inlinegraphics{figures/three.png}), that sets the value of \textit{word}. Since \textit{CLIPAgent} is watching the value of word, it is triggered to run when the user typed in ``bulldozer'' and not tied to the \textit{CLIPInputEntity} itself.

Continuing with the example in Figure \ref{fig:graph}, when the value of \textit{data} is set, the user clicks the \textit{TransferLearn} button (Figure \ref{fig:dasboard}, \inlinegraphics{figures/six.png}, but not shown in graph example) to send a message to the \textit{CNNAgent} to play its function (specified in its source code) with \textit{data} as \textit{input}. In this case, a CNN model (Xception) used for classification is fine-tuned with \textit{data}, and the weights of the model are stored as type beestar.tensor in the \textit{CNNAgent}`s \textit{output} property. A DisplayEntity of type GraphEntity (Figure \ref{fig:dasboard}, \inlinegraphics{figures/four.png}, but not shown in graph example) updates with the validation accuracy per epoch during training.

\newlength{\mintednumbersep}

\begin{listing}
\begin{minted}
[
xleftmargin=10pt,
numbersep=5pt,
frame=lines,
framesep=2mm,
baselinestretch=1.2,
fontsize=\scriptsize,
linenos,
breaklines,
]{python}
from beestar.agent_functions import clip
prompt = Entity("prompt")
input = InputEntity(name="CLIPInputEntity")
input.sets(prop="word", entities=[prompt])
agt = AgentEntity(name="CLIPAgent", func=clip)
agt.watch(prop="word", entities=[prompt])
\end{minted}
\caption{Example in \((Bee)*\) API}
\label{listing:1}
\end{listing}

\subsection{Scale and Efficiency}
Agents themselves run in parallel, supporting parallel computation, and can be automatically deployed locally as processes or pods on Kubernetes \cite{noauthor_production-grade_nodate}. Deploying agents on Kubernetes supports the cases where an agent needs a scaled environment and high bandwidth (e.g., GPUs and $\ge$  10 Gbps network). For Kubernetes, \((Bee)*\) uses a predefined base container to create an agent pod on Kubernetes, installs the needed requirements from the agent's requirements property, and launches the agent inside the container (mapping ports as needed at the cluster level). From the developer's perspective, they simply change a parameter to deploy locally or on Kubernetes and specify requirements in the agent's property, making it easy for the developer to scale their application.

\subsection{\((Bee)*\) Interface and Developer Experience}
While the graph can grow in complexity, all the specification and update logic is handled through \((Bee)*\) Interface library to support a simpler developer experience and enforce the watches and sets relationships. For declarative specification, the developer declares they would like an agent, widgets, entities, and update rules by calling methods in the \((Bee)*\) Interface. Listing \ref{listing:1} shows the sequence of calls to create a prompt Entity, an InputEntity that sets the \textit{prompt}'s \textit{word} after the user types in their keyword, and a \textit{CLIPAgent} that watches the \textit{word} or \textit{prompt}. With six lines, the \((Bee)*\) runtime is able to create a web application that takes a word from an input textbox and passes the word to a large foundational model (CLIP) so that it can run inference.

The \((Bee)*\) Interface library also handles update rules. When any property value is being updated, it has to happen through the \((Bee)*\) Interface (as shown with the boxes on the edges in Figure \ref{fig:arch}). On property change, the \((Bee)*\) Interface queries the graph for sets/watches relationships and, in turn, updates values and notifies watchers as needed. The \((Bee)*\) Interface library need not be centralized as long as the update rules are consistent across all instances.

\section{Feasibility of Developing with \((Bee)*\)}
\label{sec:Evaluation}

\begin{figure*}
    \centering
    \includegraphics[width=0.60\textwidth]{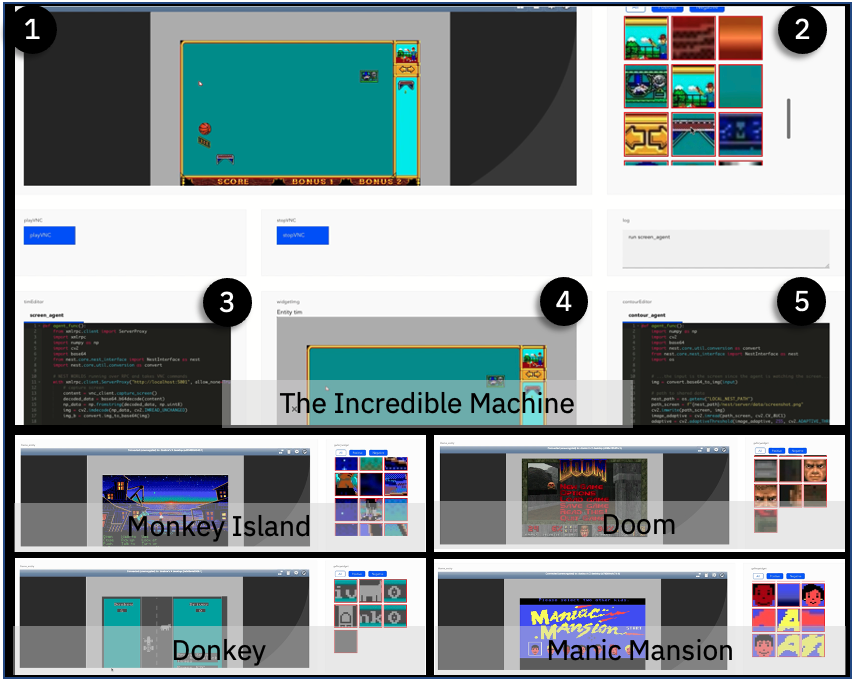}
    \caption{Same \((Bee)*\) program different games \cite{noauthor_incredible_nodate, noauthor_download_nodate-1, noauthor_doom_nodate, noauthor_donkey_nodate, noauthor_download_nodate}.}
    \label{fig:Dos games}
\end{figure*}

We  looked at how efficient the application in Figure \ref{fig:dasboard} is, to understand overhead in \((Bee)*\), and also explored a breadth of game applications that can quickly be played by agents. When running the \((Bee)*\) application in Figure \ref{fig:dasboard} on a machine with one P100 GPU on Kubernetes, we found that the \textit{CLIPAgent} was able to label 4000 construction images in 52.1s and the \textit{CNNModel} agent trained in 111.7 seconds.  Since labeling is a labor intensive tasks, the speed of the agents speaks to the feasibility of \((Bee)*\) to create programs that are fast enough to be useful.

We observed how feasible it might be to build a variety of agents working together to play arbitrary games in Figure \ref{fig:Dos games}. In Figure \ref{fig:Dos games}, \inlinegraphics{figures/one.png}, the DOS \cite{noauthor_dosbox_nodate} game called The Incredible Machine \cite{noauthor_incredible_nodate} is being played by a variety of agents. Figure \ref{fig:Dos games}, \inlinegraphics{figures/three.png} is a code editor widget for the \textit{VNCagent} that takes screen shots of the game (Figure \ref{fig:Dos games}, \inlinegraphics{figures/four.png}), and Figure \ref{fig:Dos games}, \inlinegraphics{figures/five.png} is a code editor widget for a \textit{ObjectDetectorAgent} that identifies objects from \textit{VNCagent}'s screen shot. The objects detected are set as a value for the \textit{Game} Entity, which the gallery widget (Figure \ref{fig:Dos games}, \inlinegraphics{figures/two.png}) watches and its images update as the objects update. From this \((Bee)*\) program, we were able to use the same agents and widgets to play Monkey Island \cite{noauthor_download_nodate-1}, Doom \cite{noauthor_doom_nodate}, Donkey \cite{noauthor_donkey_nodate}, and Manic Mansion \cite{noauthor_download_nodate}, by only changing the URL of where the game is being hosted in the graph. Further, the editors for the agents let the programmer tweak the code in real time for each game, showing the value meta-programming.

\section{Related Work}
\label{sec:Related_Work}
Borowski, et al. present the Varv \cite{borowski_varv_2022} declarative programming language that supports building a website live in the browser, where components live in the DOM \cite{keith_dom_2010} of the browser. Similarly, Perez De Rosso et al. present a declarative framework for websites with components called concepts \cite{perez_de_rosso_declarative_2019}. While \((Bee)*\) is declarative and supports dynamic applications, it differs in that it can declaratively integrate AI components, where the AI could be in other languages, frameworks, and infrastructure that are outside the browser, and, additionally, it can integrate agents with GUI elements in a browser. Together, these features provide a framework for integration, interaction, and scale of AI components not previously supported.

In the AI literature, using a graph to integrate agents has been demonstrated by Goertzel, et al. with the OpenCog framework \cite{goertzel_opencog_2014}, where the focus is on collaboration among agents. Our work addresses using a graph to also integrate application components (e.g., GUI components). 

Low code frameworks (e.g., Node-Red \cite{noauthor_node-red_nodate}, ConveyorAi \cite{noauthor_conveyor_nodate}, and Patterns \cite{noauthor_patterns_nodate}) use a directed graph to specify the execution and information flow of services. In contrast, we present a graph that is a shared knowledge representation, using a blackboard architecture, that is used and reflected on by agents and users for interaction and collaboration, but, further, includes components to specify a rich user interface. 
\section{Future Plans}
\label{sec:Future_Plans}

Future work includes both an in depth evaluation of \((Bee)*\) to rapidly prototype compositional AI applications and also demonstrations of using the framework to further illustrate the power of the approach of agents collaborating over the graph.

For evaluation, we will measure the speed developers can create AI applications with \((Bee)*\) versus popular approaches today. For demonstration, we will show agents searching the graph for other agents to ``outsource'' work to them, show gradients passed across agents so they learn together, and show public graphs to share agents and widgets across developers.
\section{Conclusions}
\label{sec:Conclusion}
\((Bee)*\) is a novel rapid development framework for compositional AI applications. We demonstrated how a developer can declaratively specify agents, widgets, and their relationships for rapidly integrating and scaling dynamic compositional AI applications. With the sets and watches relationships between agent and widget entities, \((Bee)*\) provides a dynamic agent and user \textbf{interaction}. For \textbf{integration} across different kinds of agents, \((Bee)*\) integrates agents and widgets at properties in the graph and not inside the agent. \((Bee)*\) supports \textbf{scale} and parallelism by running agents in parallel and on Kubernetes. 

We looked at examples to explore the feasibility of building compositional AI with \((Bee)*\). We showed CLIP and CNN models collaborating as agents to label thousands of images and train a CNN model in minutes. Further, we observed the modularity of \((Bee)*\) by showing game playing agents cooperating across five different DOS games, where only the URL of the game changed across each of the applications. 
\section{Acknowledgements}
\label{sec:Acknowledgements}
We would like to thank Grady Booch, John Cohn, Adriana Meza Soria, and Ryan Anderson for their input. We also thank the MIT-IBM Watson AI Lab for supporting this work.
\bibliographystyle{IEEEtran}
\bibliography{Main}

\end{document}